\documentclass[aps, tightenlines, letterpaper, onecolumn, superscriptaddress, notitlepage, 11pt, groupedaddress]{revtex4-1}
\usepackage{times}

\usepackage{xcolor}
\usepackage{amsmath,amsfonts,amssymb}
\usepackage{graphicx}
\usepackage[caption=false]{subfig}
\usepackage{enumerate}
\usepackage{mathrsfs}
\usepackage{enumitem}

\usepackage{tikz}
\usetikzlibrary{decorations.pathreplacing,calc}

\usepackage[pdfpagelabels,pdftex,bookmarks,breaklinks]{hyperref}
\definecolor{darkblue}{RGB}{0,0,127} 
\definecolor{darkgreen}{RGB}{0,150,0}
\hypersetup{colorlinks, linkcolor=darkblue, citecolor=darkgreen, filecolor=red, urlcolor=blue}
\hypersetup{pdfauthor={Simon Burton, Courtney G. Brell, Steven T. Flammia}}
\hypersetup{pdftitle={Classical Simulation of Quantum Error Correction in a Fibonacci Anyon Code}}

\usepackage[normalem]{ulem}

\begin{document}

\title{A Short Guide to Anyons and Modular Functors}

\author{Simon Burton}
\affiliation{Centre for Engineered Quantum Systems, School of Physics, 
The University of Sydney, Sydney, Australia}

\date{\today}

\begin{abstract}
To the working physicist, anyon theory is meant to describe
certain quasi-particle excitations occurring in two dimensional
topologically ordered systems. 
A typical calculation using this theory will involve operations
such as $\otimes$ to combine anyons, 
$F^{abc}_{d}$ to re-associate such combinations,
and $R^{ab}_c$ to commute or braid these anyons. 
Although there is a powerful string-diagram notation that greatly assists these
manipulations, we still appear to be operating on particles
arranged on a one-dimensional line, algebraically ordered from left to right.
The obvious question is, where is the other dimension?
The topological framework for considering these anyons as truly living in a
two dimensional space is known as a modular functor, or topological
quantum field theory. 
In this work we show how the apparently one-dimensional algebraic anyon theory
is secretly the theory of anyons living in a fully two-dimensional system.
The mathematical literature covering this secret is vast, 
and we try to distill this down into something more manageable.
\end{abstract}

\maketitle



\newcommand{\Eref}[1]{Eq.~(\ref{#1})}
\newcommand{\Fref}[1]{Fig.~\ref{#1}}
\newcommand{\Aref}[1]{Appendix~\ref{#1}}

\newcommand{\e}{\mathrm{e}}
\newcommand{\vac}{\mathbb{I}}

\newcommand{\ket}[1]{|{#1}\rangle}
\newcommand{\expect}[1]{\langle{#1}\rangle}
\newcommand{\bra}[1]{\langle{#1}|}
\newcommand{\ketbra}[2]{\ket{#1}\!\bra{#2}}
\newcommand{\braket}[2]{\langle{#1}|{#2}\rangle}
\newcommand{\proj}[1]{\ketbra{#1}{#1}}

\newcommand{\cggb}[1]{\textcolor{blue}{#1}}
\newcommand{\simon}[1]{\textcolor{red}{#1}}
\newcommand{\stf}[1]{\textcolor{green}{#1}}

\newcommand{\F}{\mathscr{H}} 
\newcommand{\C}{\mathbb{C}}
\newcommand{\R}{\mathbb{R}}
\newcommand{\A}{\mathcal{A}}
\newcommand{\Hom}{{\mathrm{Hom}}}

\newcommand{\subsub}[1]{{\bf #1}}

In this work we describe the theory of two-dimensional topologically ordered systems
with anyonic excitations. 
There are two main approaches to defining these,
one being more algebraic and the other more topological.

The algebraic side is known to mathematicians as the study of braided
fusion tensor categories, or more specifically, modular tensor categories.
This algebraic language appears to be more commonly used 
in the physics literature, such as the well cited Appendix E of Kitaev~\cite{Kitaev2006}.
Such algebraic calculations can be interpreted as
manipulations of string diagrams~\cite{Baez2010},
or \emph{skeins}.
These strings encode connections in the algebra, for example 
the Einstein summation convention. But they also faithfully
encode the twisting that occurs when anyons are re-ordered or
braided around each other. And these kinds of spatial relationships
are fruitfully studied using topological methods.

Working from the other direction, one starts with
a topological space (of low dimension) and attempts
to extract a combinatorial or algebraic description of how this
space can be built from joining smaller (simpler) pieces together.
These topologically rooted constructs are known as modular functors, 
or the closely related topological quantum field theories (TQFT's).

That these two approaches -- algebraic versus topological -- meet is one of the great
surprises of modern mathematics and physics.


Modular functors can be constructed from skein theory.
In the physics literature, 
this appears as skeins growing out of manifolds \cite{Pfeifer2012},
or as motivated by renormalization group considerations \cite{Levin2005}.
A further physical motivation is this: if a skein 
is supposed to correspond to the $(2+1)$-dimensional
world-lines of particles
as in the Schr\"{o}dinger picture,
modular functors would
correspond to the algebra of observables, as in
a Heisenberg picture.

In the physics literature, modular functors are
explicitly used in Refs.~\cite{Freedman2002, Freedman2002simulation}.
Also, Refs.~\cite{Beverland2014} and \cite{Kitaev2006topo}
use the language of modular functors but they call them TQFT's.
This is in fact reasonable because a modular functor can be
seen as part of a TQFT, but is quite confusing to the
novice who attempts to delve into the mathematical literature.

The definition of a modular functor appears to
be well motivated physically.
Unfortunately, there are many such definitions
in the mathematical literature 
\cite{Walker1991,Turaev1994,Bakalov2001,Tillmann1998}.
According to \cite{Bartlett2015} section 1.2 and 1.3,
there are several open questions involved in 
rigorously establishing the connection between these different axiomatizations.
In particular, what physicists call anyon theory,
and mathematicians call a modular tensor category,
has not been established to correspond exactly
(bijectively) to any of these modular functor variants. 
We try not to concern ourselves too much with these details, 
but merely note these facts as a warning to the reader
who may go searching for the ``one true formulation'' of
topological quantum field theory.

Of the many variants of modular functor found in the literature,
one important distinction to be made is
the way anyons are labeled.
In the mathematical works \cite{Turaev1994, Bakalov2001, Tillmann1998} 
we see that anyons are allowed to have superpositions
of charge states. 
However, in this work we restrict anyons to have
definite charge states, as in \cite{Walker1991,Freedman2002simulation,Beverland2014}. 
This seems to be motivated physically as such configurations
would be more stable. 

The main goal of the present work is to sketch how a
braided fusion tensor category arises from a modular functor.
In the mathematical literature,
this is covered in \cite{Turaev1994,Tillmann1998,Bakalov2001}
but as we just noted they use a different formulation for a
modular functor.




%
%

\subsection{Topological Exchange Statistics}

In this section we begin with the familiar question 
of particle exchange statistics in three dimensions, whose answer is bosons and fermions.
We then show how in restricting the particles to two dimensions many
more possibilities arise.
Our focus will be on the close connection between the algebraic and the topological viewpoints,
aiming to motivate the definition of a modular functor given in the next section.

In three spatial dimensions, the process of winding one
particle around another, a \emph{monodromy}, is topologically trivial.
This is because the path can be deformed back to the
identity; there is no obstruction:
\begin{center}
\includegraphics[]{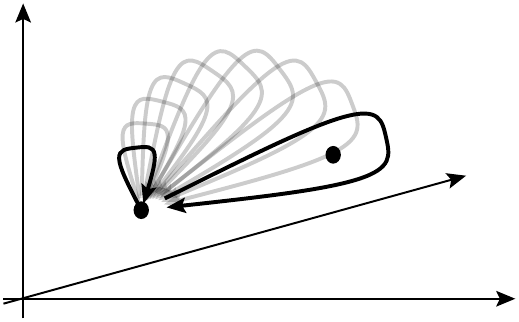}
\end{center}
The square root of this operation is a swap:
\begin{center}
\includegraphics[]{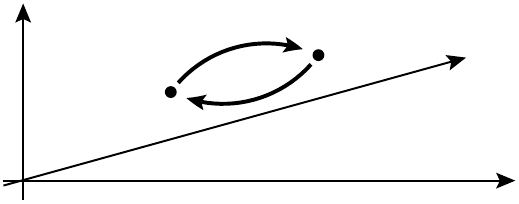}
\end{center}
For identical particles this is a \emph{symmetry} of the system. 
Continuing with this line of thought 
leads to consideration of the \emph{symmetric group}
on $n$ letters, $S_n$. 
This group is generated by the $n-1$ swap operations
$s_1,...,s_{n-1}$ that obey the relations
\begin{align*}
    s_i^2 &= 1, \\
    s_i s_j &= s_j s_i \ \ \ \mbox{for}\ \ |i-j|>1,\\
    s_i s_{i+1} s_i &= s_{i+1} s_i s_{i+1} \ \ \ \mbox{for}\ \ 1\le i \le n-2.
\end{align*}
Writing the Hilbert space of the system as $V$ 
we would then expect $S_n$ to act on this space via unitary transformations $U(V):$
$$
    \F : S_n \to U(V).
$$
This is the first example of the kind of functor we will be talking about. 
In this case it is a group representation; $\F$ is a homomorphism between
two groups.

When the above monodromy is constrained to two dimensions
we can no longer deform this process to the identity:
\begin{center}
\includegraphics[]{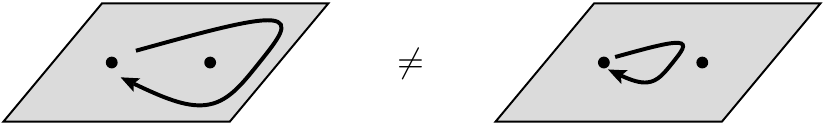}
\end{center}
and so in two dimensions we cannot expect a swap to square to the identity.
To see this more clearly,
we must examine the entire (2+1)-dimensional \emph{world-lines} of these particles.
For an example, here we show the world-lines of three particles undergoing an exchange and then
returning to their original positions:
\begin{center}
\includegraphics[]{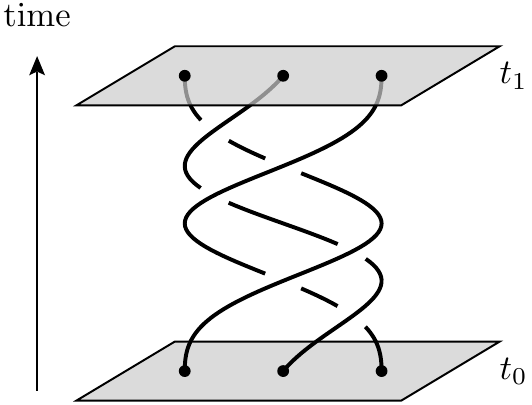}
\end{center}
Note that if we allow the particles to move in one extra
dimension then we can untangle these braided world lines.
The question is now, what is the group that acts on the
state space from $t_0$ to $t_1$?
Examining the structure of these processes more closely,
we see that we can compose them by sequentially performing
two such braids, one after the other:
\begin{center}
\includegraphics[]{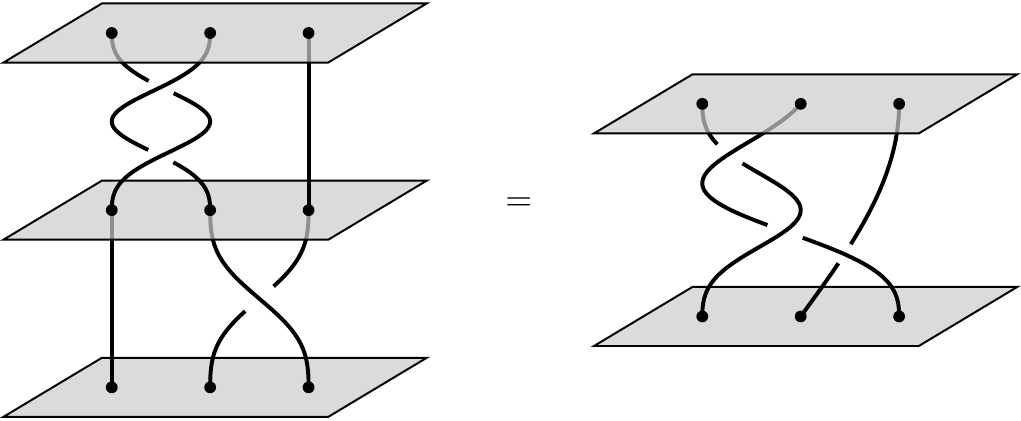}
\end{center}
And, by ``reversing time'', we can undo the effect of
any braid:
\begin{center}
\includegraphics[]{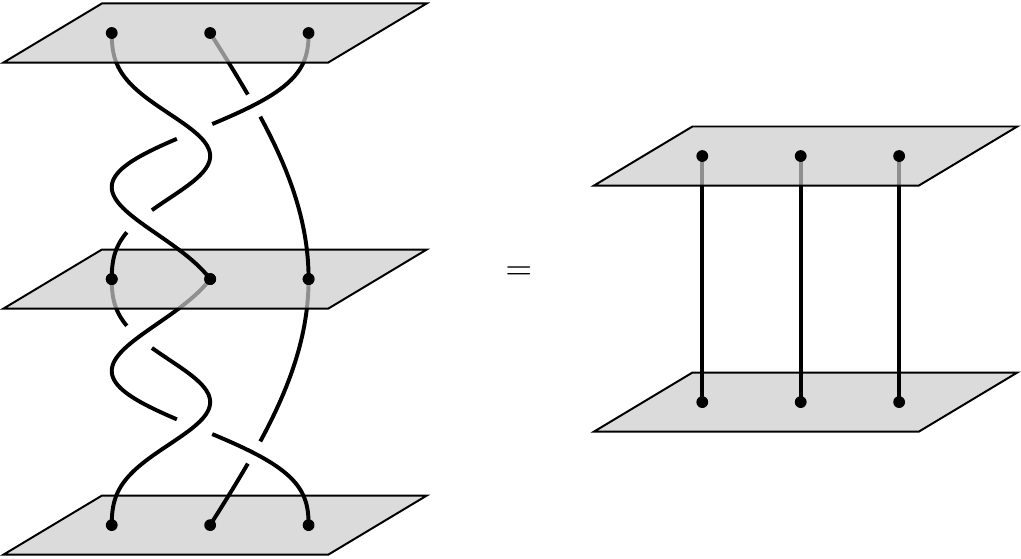}
\end{center}
This shows that these processes do form a group,
known as the \emph{braid group.}
For $n$ particle world-lines we denote this group as $B_n.$
For identical particles this group acts as symmetries of the
state space:
$$
    \F : B_n \to U(V).
$$

What we have given is a topological description of the group $B_n$.
More formally, we can describe these braid world lines as
paths in the \emph{configuration space} of $n$ points.
\newcommand{\Conf}{\mathcal{C}}
\newcommand{\UConf}{\mathcal{UC}}
This space is defined as the product of
a two dimensional space $M$ for each point, minus the subspace where points overlap:
$$
    \Conf_n = \bigl( \prod_{1}^{n} M \bigr) - \Delta, \ \ 
    \Delta = \{(x_1, ..., x_n) | x_i = x_j \ \ \mbox{for some} \ \ i\ne j \}.
$$
Because we are considering identical particles (so far)
we use the \emph{unlabelled configuration space}
$\UConf_n$ which is the quotient of $\Conf_n$ by the natural action of the
permutation group $S_n:$
$$
    \UConf_n = \Conf_n / S_n.
$$
The \emph{geometric braid group} as we have so far informally described it 
can now be rigorously defined as the fundamental group:
$$
    B_n = \pi_1 ( \UConf_n, x )
$$
where $x$ is some reference configuration in $\UConf_n.$
See Ref.~\cite{Ghrist2014} for further discussion.

This group also has a purely algebraic description via
generators and relations, as was shown by Artin in 1947, \cite{Artin1947,Birman1974}.
In this description, 
$B_n$ is generated by $n-1$ elements $\sigma_1,..,\sigma_{n-1}$ that satisfy
the following relations:
\begin{align*}
    \sigma_i \sigma_j &= \sigma_j \sigma_i \ \ \ \mbox{for}\ \ |i-j|>1,\\
    \sigma_i \sigma_{i+1} \sigma_i &= \sigma_{i+1} \sigma_i \sigma_{i+1} \ \ \ \mbox{for}\ \ 1\le i \le n-2.
\end{align*}
These relations are the same as for $S_n$ above, except 
we do not require $\sigma_i^2 = 1.$

It is easy to see that the geometric braid group satisfies these
relations. Here we show the braids corresponding to the generators $\sigma_i:$
\begin{center}
\includegraphics[]{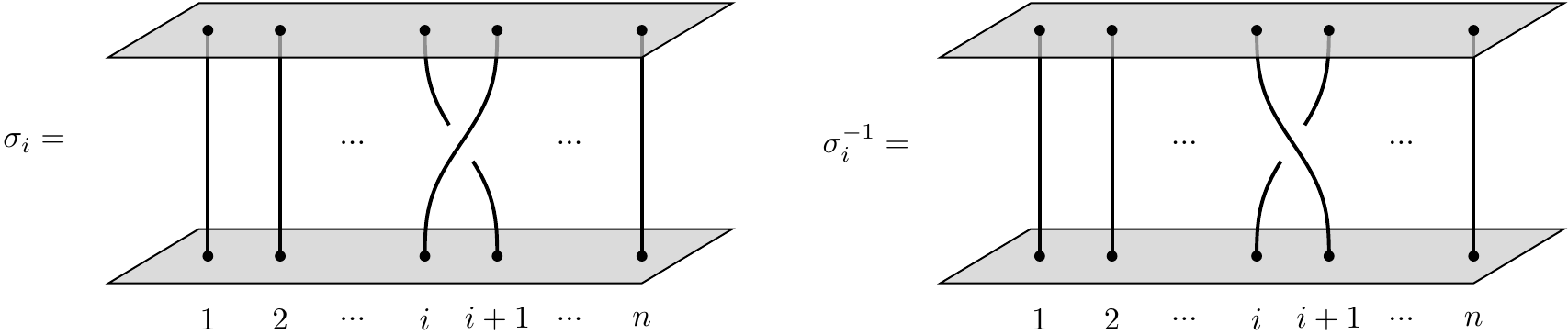}
\end{center}
So that 
$\sigma_i \sigma_j = \sigma_j \sigma_i \ \ \ \mbox{for}\ \ |i-j|>1$
because the two braids are operating on disjoint world-lines.
The second relation is also easy to see:
\begin{center}
\includegraphics[]{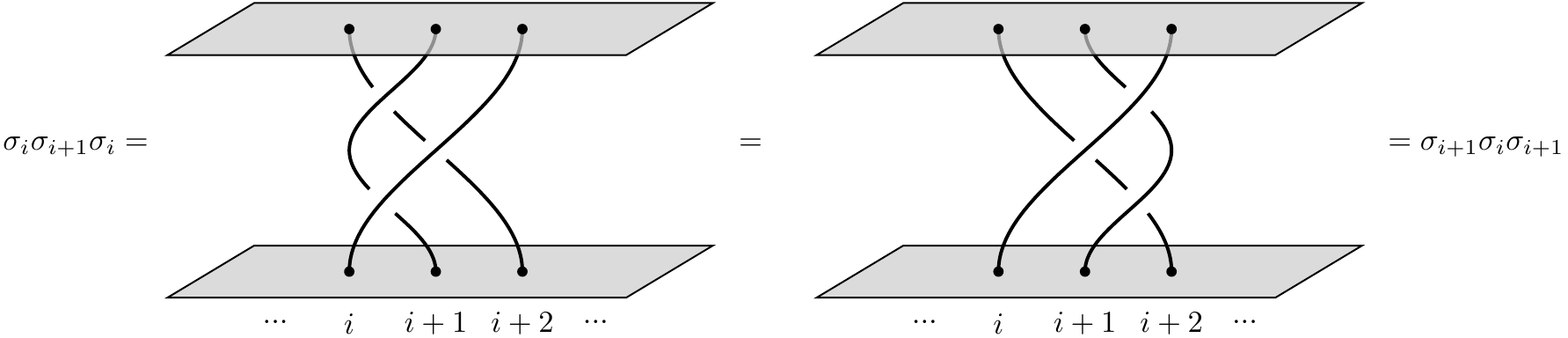}
\end{center}

There is another important geometric representation of
the braid group which is purely two-dimensional.
We pick an $n$-point finite subset $Q_n \subset M$
and consider diffeomorphisms 
$f : M \to M$
that map $Q_n$ to itself.
The \emph{mapping class group} of $M$ relative to $Q_n$
is the set of all such diffeomorphisms up to an equivalence relation $\sim_{\mbox{iso}}$:
$$
    MCG(M, Q_n) = \{ f : M \to M \ \ \mbox{such that}\ \ f(Q_n)=Q_n \} / \sim_{\mbox{iso}}
$$
The equivalence relation $\sim_{\mbox{iso}}$ is called \emph{isotopy}
which allows for any continuous deformation of $f:M\to M$ that
fixes each point in $Q_n.$
Each generator $\sigma_i$ of the braid group is found in $MCG(M, Q_n)$
as a \emph{half-twist} that swaps two points $i, i+1 \in Q_n$:
\begin{center}
\includegraphics[width=1.0\columnwidth]{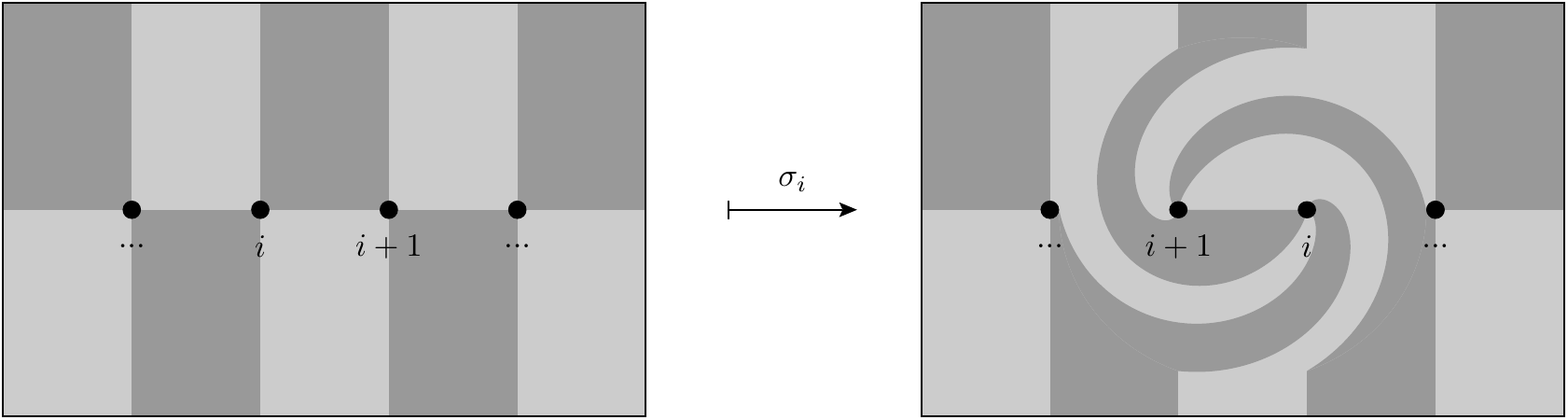}
\end{center}
In order to show
the action of
this half-twist we have 
decorated the manifold with
a a checkered pattern,
but there is a more important
object that lives on the manifold
itself.
An \emph{observable} is a simple
closed curve in $M$ that does
not intersect $Q_n.$
Such a closed curve is called an
observable because these will be
associated to measurements
of the total anyonic charge on the
interior of the curve.
The importance of understanding the
braid group as identical to the mapping class group
is now manifest:
whereas geometric braids act on states
as in a Schr\"{o}dinger picture,
elements of the mapping class group
act on the observables as in
a Heisenberg picture.
\begin{center}
\includegraphics[]{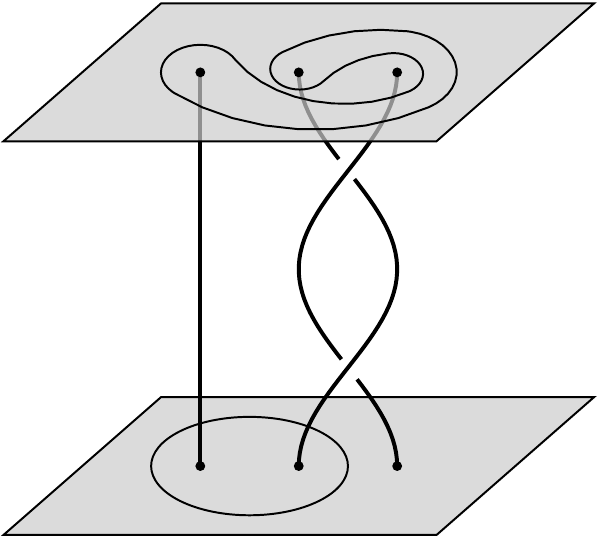}
\end{center}
This diagram should give the reader some idea as
to why these two definitions of the braid group are
equivalent, but the actual proof of this is somewhat involved.
We cite Ref.~\cite{Kassel2010} for an excellent contemporary
account that fills in these gaps.


The definition given above for the geometric braid group
and the mapping class group make sense for any two dimensional
manifold $M$ but for concreteness we consider $M$ to be
a flat disc. 
The corresponding algebraic definition of the braid group
will in general be altered depending on the underlying manifold
$M.$

So far we have been studying the exchange statistics
for $n$ identical particles. Without this restriction, one needs to
constrain the allowed exchange processes so as to preserve particle type.
For example, if all particles are different we would use the
\emph{pure braid group} $PB_n.$
The geometric description of
this group is as the fundamental group of the labelled
configuration space
$$
    PB_n = \pi_1 ( \Conf_n, x ).
$$
Loops in $\Conf_n$ correspond to braids where each world-line returns
to the point it started from.
The algebraic description of this group is somewhat complicated and
we omit this.
In terms of the mapping class group, we can describe $PB_n$
as the \emph{pure mapping class group}:
$$
    PMCG(M, Q_n) = \{ f : M \to M \ \ \mbox{such that}\ \ f(x)=x \ \ \mbox{for}\ x\in Q_n\} / \sim_{\mbox{iso}}.
$$

One further complication arises when particles have a rotational
degree of freedom: ie., they can be rotated by $2\pi$ in-place and
this effects the state of the system.
To capture this action, we use the \emph{framed braid group} $FB_n.$
This group can be presented algebraically using the same generators
and relations as for the braid group $B_n,$ along with
``twist'' generators $\theta_i$ for $i=1,..,n.$
These must satisfy the further relations
\begin{align*}
    \theta_i \theta_j &= \theta_j \theta_i \\
    \theta_i\sigma_j &= \sigma_j\theta_i 
    \ \ \mbox{if}\ \ i<j \ \ \mbox{or}\ \ i\ge j+2\\
    \theta_{i+1}\sigma_i &= \sigma_i\theta_i \\
    \theta_{i}\sigma_i &= \sigma_i\theta_{i+1}.
\end{align*}

Here we show four 
geometric approaches to representing a twist.
Any person that has struggled to untangle their headphone cable
will immediately see what is going on here.
\begin{center}
\includegraphics[]{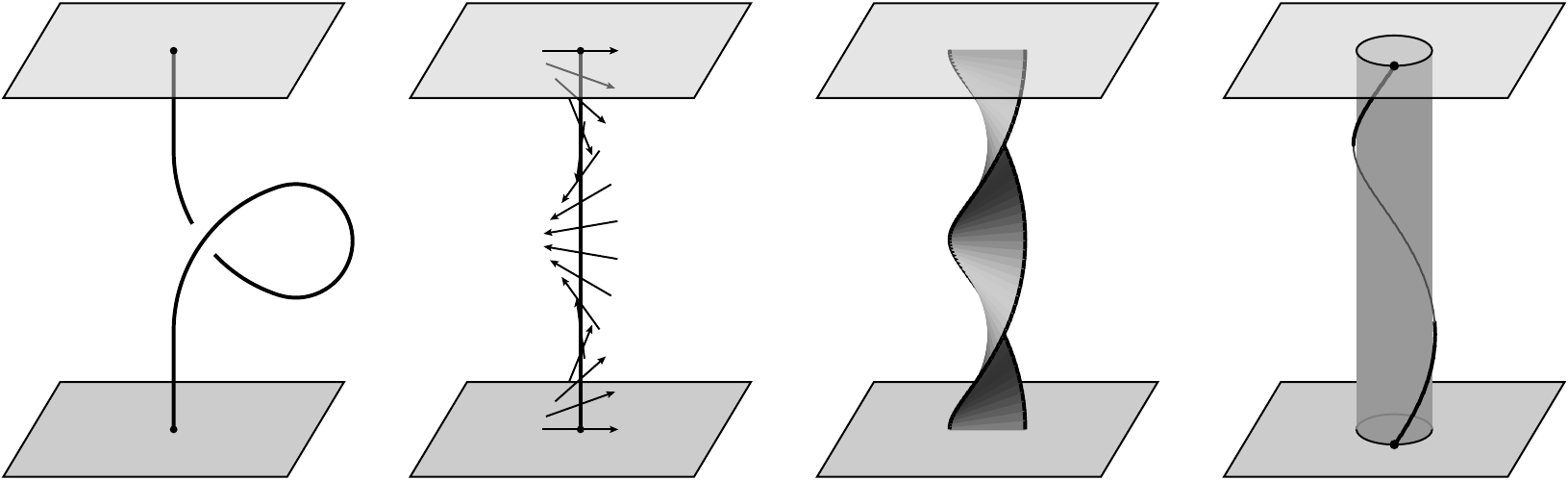}
\end{center}
On the left we have a loop; it is not a braid because it
travels backwards in time.
If we pull on this loop to make it straight, we introduce a twist.
This is shown in the next figure, 
where we show a \emph{framing} which is a non-degenerate 
vector field along the world-lines of a braid.
The initial and final vectors in the vector field must be the same.
By non-degenerate we mean that the vector field is everywhere
non-zero and non-tangent to the world-line.
In the next figure we show a \emph{ribbon}:
instead of point particles we have short one-dimensional curves in $M$.
On the right the particle is represented as a boundary
component (a hole) of $M$ with a distinguished point.
In this picture the world-line looks like a tube.

All these representations of twists carry
essentially the same information.
In the sequel we will stick
to thinking of particles as boundary
components because this fits well
with the way we are formulating 
observables as simple closed curves in $M$.

At this point in the narrative we are 
close to our next destination.
All of these considerations,
of framed or unframed, 
labelled or not,
with possibly different underlying manifolds,
together with observables,
is mean to
be captured by the formalism of a modular functor which we turn to next.

%
%

\subsection{Modular Functors}

We list the axioms for a 
\emph{2-dimensional unitary topological modular functor}.

For our purposes a \emph{surface} will be a compact
oriented $2$-dimensional differentiable manifold with boundary. 
We will not require surfaces to be connected.
By a \emph{hole} of $M$ we mean a connected component of its boundary. 
Each hole will inherit the manifold orientation,
contain a distinguished \emph{base point},
and be labeled with
an element from a fixed finite set $\A.$
This is the set of ``anyon labels'', and comes
equipped with a vacuum element $\vac$ 
and an involution $\ \widehat{}\ $ such that $\widehat{\vac}=\vac.$
The involution maps an anyon label to the ``antiparticle'' label.

We will mostly be concerned with planar such
surfaces, that is, a \emph{disc} with holes removed from the
interior.
Such surfaces will be given a clockwise 
orientation, which induces a counterclockwise orientation on
any \emph{interior} hole and a clockwise orientation
on the \emph{exterior} hole.
\begin{center}
\includegraphics[]{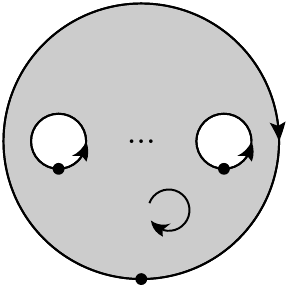}
\end{center}
Although it helps to draw such a surface as a disc with holes,
we stress that there is no real distinction to be made between
interior holes and the exterior hole.
That is, 
a disc with $n$ interior holes is (equivalent to) a sphere with $n+1$ holes.
We merely take advantage of the fact that a sphere with at least one hole
can be flattened onto the page by ``choosing'' one of the holes to
serve as the exterior hole.

By a \emph{map of surfaces} $f:M\to N$ we mean 
a diffeomorphism that preserves
manifold orientation, hole labels, and base points.
Note that we also deal with maps of various other objects
(vector spaces, sets, etc.)
but a map of surfaces will have these specific
requirements.

Two maps of surfaces $f:M\to N$ and $f':M\to N$
will be called \emph{isotopic}
when one is a continuous deformation of the other.
In detail,
we have a continuously
parametrized family of maps $f_t:M\to N$ for $t\in [0, 1]$
such that $f_0=f,\ f_1=f'$
and the restriction of $f_t$ to the 
set of marked points $X\subset\partial M$ is constant:
$f_t|_{X} = f|_{X}$ for $t\in [0, 1].$
Such a family $\{f_t\}_{t\in [0,1]}$ is called an \emph{isotopy}
of $f.$
This is an equivalence relation on maps $M\to N$, and the equivalence
class of $f$ under isotopy is called the \emph{isotopy class} of $f.$
The weaker notion of homotopy of maps will not be used
here, but for the maps we use
it turns out that homotopy is equivalent to isotopy.
Furthermore, we can weaken the requirement that maps be differentiable,
because every continuous map $f:M\to N$ is (continuously)
isotopic to a differentiable map~\cite{Farb2011}.

A \emph{modular functor} $\F$ associates to every
surface $M$ a finite dimensional complex
vector space $\F(M)$, called the \emph{fusion space} of $M$.
For each map $f:M\to N$
the modular functor associates a
unitary transformation $\F(f) : \F(M)\to\F(N)$
that only depends on the isotopy class of $f.$
Functoriality requires that $\F$
respect composition of maps.

We have the following axioms for $\F.$

{\bf Unit axioms.}
The fusion space of an empty surface is one dimensional, 
$\F(\phi) \cong \C.$ 
For $M_a$ a disc with boundary label $a$ we have
$\F(M_\vac)\cong \C$ and 
$\F(M_a)\cong 0$ for $a\ne \vac.$
For an annulus $M_{a,b}$ with boundary labels
$a, b$ we have
$\F(M_{a,\widehat{a}})\cong \C$ and 
$\F(M_{a,b})\cong 0$ for $a\ne \widehat{b}.$

{\bf Monoidal axiom.}
The disjoint union of two surfaces $M$ and $N$ 
is associated with
the tensor product of fusion spaces:
$$
    \F(M\amalg N) \cong \F(M)\otimes \F(N).
$$
This is natural from the point of view of quantum
physics, where the Hilbert space of two disjoint
systems is the tensor product of the space for
each system.

{\bf Gluing axiom.}
Denote a surface $M$ with (at least) two holes
labeled $a, b$ as $M_{a,b}$. 
If we constrain $b=\widehat{a}$ 
then we may \emph{glue} these two holes together
to form a new surface $N$.
To construct $N$ we choose a diffeomorphism from one
hole to the other that maps base point to base point 
and reverses orientation.
Identifying the two holes along this diffeomorphism
gives the glued surface $N$.
(There is a slight technicality in ensuring that $N$ is
then differentiable, but we will gloss over this detail.)
The image of these holes in $N$ we call a \emph{seam}.
The fusion space of $M_{a,\widehat{a}}$ then embeds unitarily in the fusion
space of $N$.
Moreover, there is an isomorphism called
a \emph{gluing map}:
$$
    \bigoplus_{a\in\A} \F(M_{a,\widehat{a}}) \xrightarrow{\cong} \F(N)
$$
and this isomorphism depends only on the isotopy class of 
the seam in $N.$

{\bf Unitarity axiom.}
Reversing the orientation of the surface $M$
to form $\overline{M}$ we get the dual of the fusion space:
$\F(\overline{M})=\F(M)^*.$

\subsub{Compatibility axioms.}
Loosely put, we require that the above operations play nicely together,
and commute with maps of surfaces.
For example,
a sequence of gluing operations 
applied to a surface can
be performed regardless of the order (gluing is associative)
and we require the various gluing maps for these operations to similarly agree.
For another example, we require $\F$ to respect that gluing
commutes with disjoint union.

\subsub{Observables.}
The seam along which gluing occurs can be associated
with an observable as follows.
We take a gluing map $g$ and projectors $P_a$ onto the
summands in the above direct sum:
$$
    P_a: \bigoplus_{b\in\A} \F(M_{b,\widehat{b}}) \to \F(M_{a,\widehat{a}}).
$$
then the observable will be the set of operators 
$\{P_a g^{-1}\}_{a\in\A}.$
For each $a\in A$
we call the image of $P_a$ 
a \emph{charge sector} for that observable.
Note that in the glued surface $N$ the seam has no
preferred orientation (or base point).
If we choose an orientation 
for the seam this corresponds to choosing one of the
two boundary components in the original surface $M_{a,\widehat{a}}.$
If these boundary components come from disconnected components of 
$M_{a,\widehat{a}}$ the seam cuts $N$ into two pieces and the
orientation chooses an \emph{interior:}
following the orientation around the seam presents
the interior to the left.
We intentionally confuse the distinction between an
observable as a set of operators,
and the associated seam along which gluing occurs.

\subsub{Consequences of axioms.}
A common operation is to glue two separate surfaces.
We can do this by first taking disjoint union (tensoring
the fusion spaces)
and then gluing.
Here we show this process applied to two surfaces $M$ and $N$. 
\begin{center}
\includegraphics[]{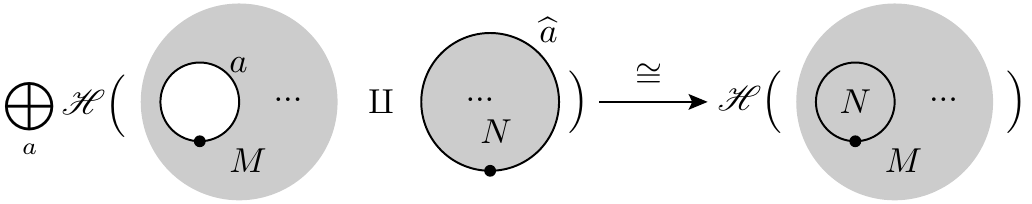}
\end{center}
We display the surface $N$ with the $\widehat{a}$
boundary on the outside, 
to show more clearly how $N$ fits into $M$.
In the glued surface we
indicate the placement of $M$ and $N$ and the seam
along which gluing occurred,
as well as the identification of base points.

We note two other consequences of the axioms.
A hole of $M$ labeled with $\vac$ can be replaced with a disk (by gluing)
and this does not change the fusion space of $M.$
That is, a hole that carries no charge can be ``filled-in''.
And, the dimensionality of the fusion space of a torus is the
cardinality of $\A.$
This can be seen by gluing one end
of an annulus (cylinder) to the other.

\subsub{Fusion.}
When a surface can be presented as the gluing of
two separate surfaces, we have projectors onto the
fusion space of either glued surface:
\begin{center}
\includegraphics[]{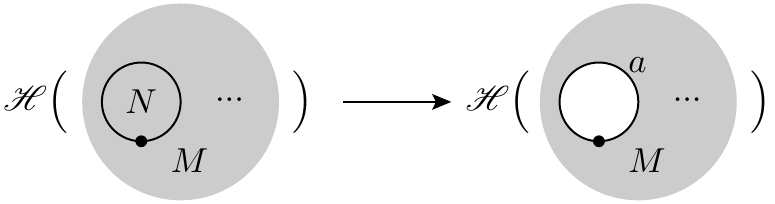}
\end{center}
In this case, we define 
the operation of \emph{fusion} to replace the interior
of an observable by a single hole.
This is an operation on the manifold itself,
and we will only do this when the interior piece is a
disc with zero or more holes.

%

\subsub{F-move.}
The fusion space of the disc and annulus are
specified by the axioms, and we define the fusion
space of the disc with two holes, or \emph{pair-of-pants} as:
\begin{center}
\includegraphics[]{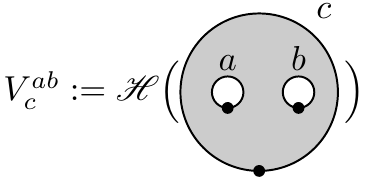}
\end{center}

The \emph{$F$-move}
is constructed from two applications of a
gluing map (one in reverse) as the following commutative diagram shows:
\begin{center}
\includegraphics[]{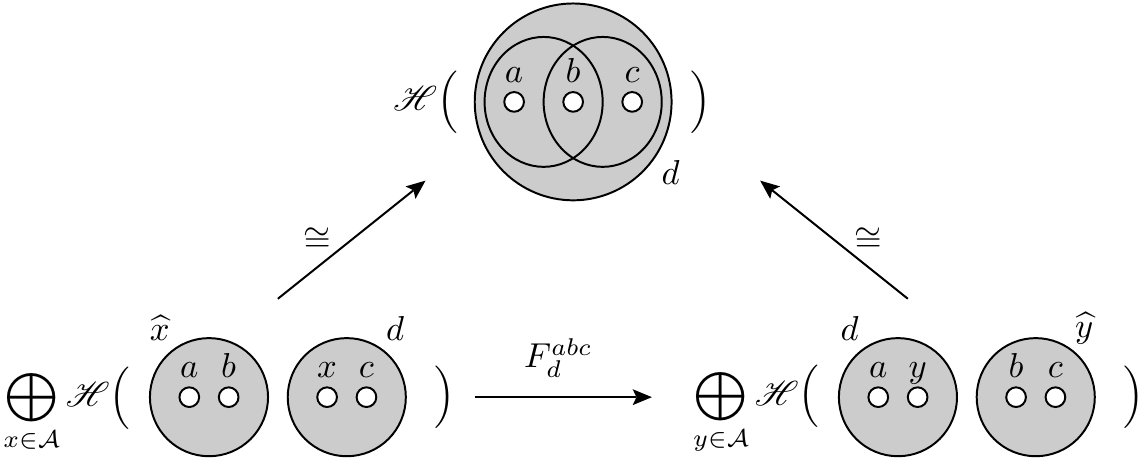}
\end{center}
Here we have a surface $M^{abc}_d$ with 
four labeled boundary components, as well as two separate
ways of gluing pairs-of-pants to get $M^{abc}_d.$
We can also write this out in terms of the fusion spaces of pair-of-pants:
$$
    F^{abc}_d : \bigoplus_{x\in\A} V^{ab}_{\widehat{x}}\otimes V^{xc}_d 
        \to \bigoplus_{y\in\A} V^{ay}_d\otimes V^{bc}_{\widehat{y}}.
$$
By using gluing, and summing over charge sectors,
we can extend this operation 
to apply where any of the boundary components $a, b, c$ or $d$ are merely 
seams in a larger manifold.


\subsub{POP decomposition.}
Given a manifold $M$ which is
a disc with two or more holes, 
we show how to present $M$ as the gluing of various
pair-of-pants. Such an arrangement will be termed a \emph{POP decomposition.}
(We refer to Ref.~\cite{Ivanov2001} for more details on this construction,
and Ref.~\cite{Ghrist2014} for a leisurely description of Morse theory.)
This will yield a decomposition of $\F(M)$ into a direct
sum of fusion spaces of pair-of-pants. 
The key idea is to choose a ``height'' function $h:M\to\R$
with some specific properties that allow us to
cut the manifold up along level sets of $h.$
First, we need that critical points of $h$ are isolated.
This is the defining condition for $h$ to be a \emph{Morse function}.
Also, we need that $h$ is constant on $\partial M,$
and the values of $h$ at different critical points are distinct.
Now choose a sequence of non-critical values $a_1<a_2<...<a_n$ in $\R$
such that every interval $[a_{i-1}, a_i]$ 
contains exactly one critical value of $h$ and the image
of $h$ lies within $[a_1, a_n].$
Each component of $h^{-1}([a_{i-1}, a_i])$ is then either
an annulus, a disc, or a pair-of-pants depending on the index of (any)
critical point it contains.
We then re-glue any annuli or discs until there are no more of these
and we have only pair-of-pants.
\begin{center}
\includegraphics[]{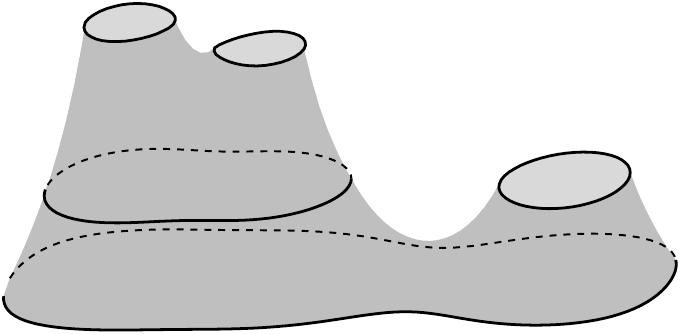}\ \ \ \ \ \ \ \ \includegraphics[]{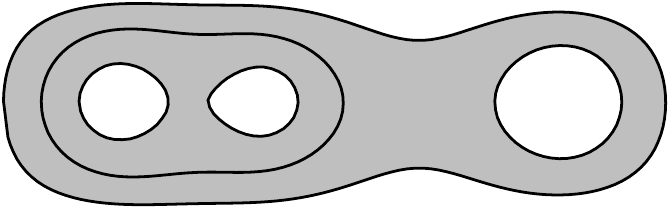}
\end{center}

Clearly such a POP decomposition is not unique,
and the goal here is to understand how to switch between decompositions,
and in particular, given an observable $\gamma$,
and a given POP decomposition, find a sequence
of ``moves'' such that $\gamma$ is in the resulting POP
decomposition:
\begin{center}
\includegraphics[]{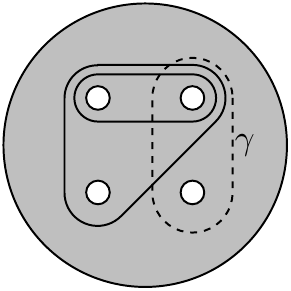}
\end{center}

One way to achieve this is via Cerf theory, which is the theory
of how one may deform Morse functions into other Morse functions
and the kind of transitions involved in their critical point structure.
This was the approach used in Ref.~\cite{Freedman2002simulation}.
In this work we use a simpler method,
which is essentially the same as {skein theory.}
This is the {refactoring theorem} that we describe below.

\subsub{Dehn twist.}
Consider a surface $M_{a,\widehat{a}}$
with two boundary components, $a$ and $\widehat{a}.$
Let $f$ be a map $M_{a,\widehat{a}}\to M_{a,\widehat{a}}$
which performs a clockwise $2\pi$ full-twist or \emph{Dehn twist}.
Here we show the action of $f$ by highlighting the
equator of the annulus:
\begin{center}
\includegraphics[]{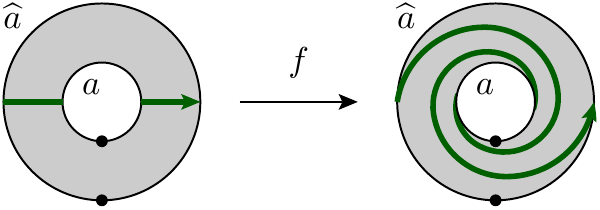}
\end{center}
We define the 
induced map on fusion spaces as $\theta_a := \F(f).$
Because $\F(M_{a,\widehat{a}})$ is one-dimensional
this will be multiplication by a complex number which we also
write as $\theta_a.$

If we now take $M$ to be an arbitrary surface, and $\gamma$ an
observable on $M,$ we can consider a neighbourhood of $\gamma$
which will be an annulus, and perform a Dehn twist there,
which we denote as $f_\gamma : M\to M.$
Writing $M$ as a gluing along $\gamma$ of another
manifold $N_{a,\widehat{a}},$
the action of $\F(f_\gamma)$ will decompose as a direct sum over charge 
sectors:
$$
\bigoplus_{a\in\A} \theta_a \F(N_{a,\widehat{a}}).
$$

\newcommand{\StdM}{M^{a_1...a_n}_{b}}

\subsub{Standard surfaces.}
For each $n=0,1,...$ and every ordered sequence of
anyon labels $a_1,...,a_n,b$ we choose a \emph{standard surface}.
This is a surface with
$n+1$ boundary components labeled $a_1,...,a_n,{b}$ 
which we denote $\StdM.$ 

For concreteness
we define this surface using the following 
expression for a closed disc less $n$ open discs:
$$
\Bigl\{(x,y)\in\R^2 \ \mathrm{st.}\  \Bigl|(x,y)-\Bigl(\frac{1}{2},0\Bigr)\Bigr|\le \frac{1}{2}\Bigr\} -
\Bigl\{(x,y)\in\R^2 \ \mathrm{st.}\  \Bigl|(x,y)-\Bigl(\frac{2i-1}{2n},0\Bigr)\Bigr| < \frac{1}{4n}\Bigr\}_{i=1,...n}.
$$
We then label the interior holes $a_1,...,a_n$ in order of
increasing $x$ coordinate, and the exterior hole is labeled $b$.
The base points are placed in the direction of negative $y$ coordinate.
As a notational convenience we highlight the \emph{equator}
of the surface 
which is the intersection of
the $x$ axis with the surface:
\begin{center}
\includegraphics[]{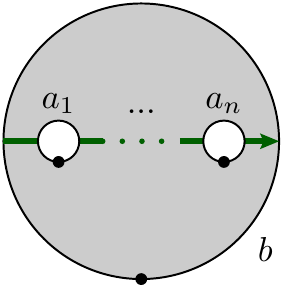}
\end{center}

Each standard surface comes with a collection of
\emph{standard POP decompositions}:
these will be POP decompositions where we require each
observable to cross the
equator twice and have counterclockwise orientation.
Up to isotopy, a given standard surface will have only finitely many of these.
On a standard surface with three
interior holes there are two standard POP decompositions,
and one $F$-move that relates these.
On a standard surface with four interior holes there are five 
standard POP decompositions and five $F$-moves that relate these.
In this case the $F$-moves themselves satisfy an equation
that is an immediate consequence of the way we have defined $F$-moves.
This is known as the pentagon equation, which we depict as
the following commutative diagram:
\begin{center}
\includegraphics[]{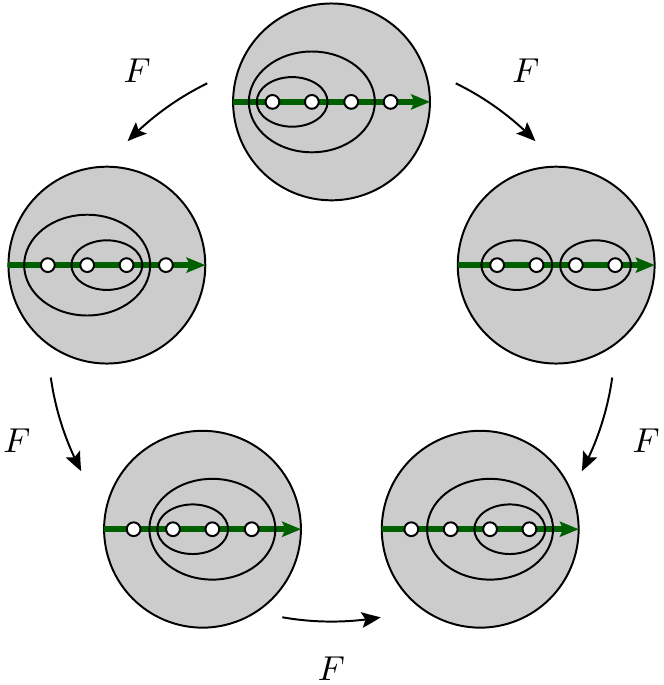}
\end{center}
We are now starting to confuse the notation 
for the topological space $M$ and the fusion space $\F(M).$
Also, each of these $F$-moves is referring to an isomorphism 
that is block decomposed according to the charge sectors 
of the indicated observables.

We define the vector spaces 
$ V^{a_1...a_n}_{b} := \F(\StdM).$ 
We now choose a basis for each of the
$V^{a_1a_2}_{b}$ to be $\{v^{a_1a_2}_{b,\mu}\}_{\mu}.$
For every standard POP decomposition of $\StdM,$
we get a decomposition of 
$V^{a_1...a_n}_{b}$ into direct sums of various 
$V^{a'_1a'_2}_{b'}.$ 
This then gives a \emph{standard basis} of $V^{a_1...a_n}_{b}$
relative to this standard POP decomposition
using the corresponding  $\{v^{a'_1a'_2}_{b',\mu}\}_{\mu}$
for each of the $V^{a'_1a'_2}_{b'}.$ 

Note that for any standard surface $\StdM$ 
and any choice of $k$ contiguous holes $a_j,...,a_{j+k-1}$
we can find an observable that encloses exactly these holes
in at least one of the standard POP decompositions of $\StdM.$ 

We next show how to glue the exterior hole of
a standard surface $M$
to an interior hole of another standard surface $N.$
To do this within $\R^2$ we rescale and translate the two surfaces so
that the exterior hole of $M$ coincides with the interior hole of $N.$
At this point the union is not in general going to
produce another standard surface,
and so we remedy this by applying any isotopy within $\R^2$
that fixes the $x$-axis while moving the surface to a standard surface.

\subsub{Curve diagram.}
We next study maps from standard surfaces to arbitrary surfaces.
The reader should think of this as akin to choosing a basis
for a vector space.
The set of all maps from $\StdM$ 
to a surface $N$ will be denoted as $\Hom(\StdM, N).$ 
We call each such map a \emph{curve diagram},
or more specifically, a curve diagram on $N$.
The reason for this terminology is that we can reconstruct (up to isotopy)
any map $f:\StdM\to N$ from
the restriction of $f$ to the equator of $\StdM.$
In other words, we can uniquely specify any map
$f:\StdM\to N$ by indicating the action of
$f$ on the equator.
(This reconstruction works because a simple closed curve on a sphere cuts 
the sphere into two discs.)
We take full advantage of this
fact in our notation: any figure of a surface
with a ``green line'' drawn therein is actually notating a 
diffeomorphism, not a surface!

Any curve diagram will act on a standard POP decomposition
of $\StdM$ sending it to a POP decomposition of $N.$
Surprisingly, the converse of this statement also holds:
any POP decomposition of $N$ (a disc with holes)
comes from some curve diagram
acting on a standard POP decomposition.
The proof of this is constructive, and we call this
the refactoring theorem below. 

The operation of 
gluing of surfaces can be extended to gluing of curve diagrams as
long as we are careful with the way we identify along the seam:
the identification map needs to respect the equator of the curve diagram. 
(Keep in mind that a curve diagram is really a map of surfaces,
and so gluing two such maps involves two separate gluing operations.)


We note in passing two connections to the mathematical literature.
Such curve diagrams have been 
used in the study of braid groups \cite{Dehornoy2002}, and this
is where the name comes from, although our
curve diagrams respect the base points and so could be further
qualified as ``framed'' curve diagrams.
And, we note the similarity of curve diagrams and associated
modular functor 
to the definition of a planar algebra \cite{Jones1999},
the main difference being that planar algebras allow for 
not just two but any even number of curve intersections at each hole.

\subsub{Z-move.}
We let $z$ be a diffeomorphism of standard surfaces 
$z : M^{a_1...a_n}_{b} \to  M^{a_2...a_n{b}}_{{a_1}}$
that preserves the equator.
(Considering the standard surface as a sphere with holes
placed uniformly around a great circle, $z$ is seen to
be a ``rotation''.)
This acts by pre-composition to
send a curve diagram $f\in \Hom(\StdM, N)$ to 
$fz \in \Hom(M^{a_2...a_n{b}}_{{a_1}}, N).$
This we call a $Z$-move of $f.$

In this way, any curve diagram can be seen as another
curve diagram that has a cyclic permutation of the labels of
the underlying standard surface.

\subsub{R-move.}
Given anyon labels $a, b$ and $c$, and arbitrary
surface $N,$
we now define the following map of curve diagrams on $N:$
$$R^{ab}_c: \Hom(M^{ab}_c, N)\to \Hom(M^{ba}_c, N).$$
This map works by taking a curve diagram 
$f:M^{ab}_c\to N$ to the composition $f\sigma$ where
$\sigma:M^{ba}_c\to M^{ab}_c$
is a counterclockwise ``half-twist'' map that exchanges
the $a$ and $b$ holes.
Here we show the action of $R^{ab}_c$ on one particular curve diagram:
\begin{center}
\includegraphics[]{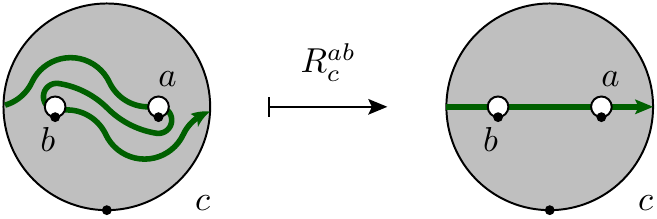}
\end{center}
Such an application of $R^{ab}_c$ to a particular
curve diagram we call an \emph{R-move.}
As noted above, a curve diagram serves to pick out
a basis for the fusion space, and the point of this $R$-move is
to switch between different curve diagrams for the same surface.
This is highlighted to draw the readers attention to
the fact that the $R$-move does not swap the labels on the holes:
the surface itself stays the same.

As we extended Dehn twists under gluing,
and $F$-moves under gluing,
we also do this for $R$-moves.


\subsub{Skeins.}
The previous figure can be seen as a ``top-down'' view of the
following three dimensional arrangement:
\begin{center}
\includegraphics[]{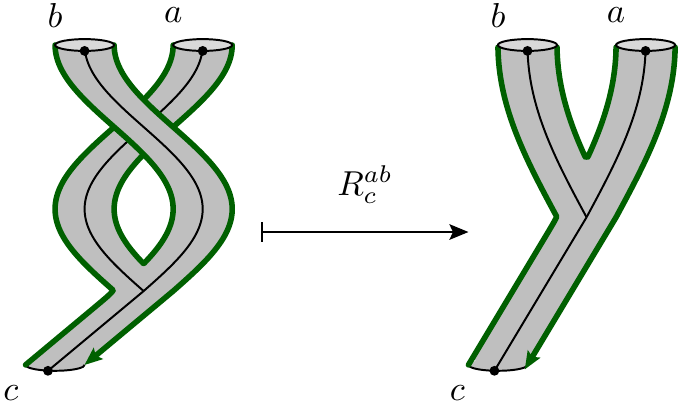}
\end{center}
This figure is intended to be topologically
the same as the previous flat figure,
with the addition of a third dimension,
and the $c$ boundary has been shrunk.
Also note thin black lines
connecting the base points.
The black lines do not add any extra structure, they can
be seen as a part of the hexagon cut out by
the green line and boundary components.
But notice this: the black lines are ``framed'' by the green lines. 
These are the ribbons used in skein theory!
Note that all the holes are created equal: there is
no distinction between ``input'' holes and ``output'' holes
(as there is with cobordisms or string diagrams.)


\subsub{Hexagon equation.}
For a surface with three interior holes there are
infinitely many POP decompositions (up to isotopy).
These are all made by choosing an observable 
that encloses two holes.
The $F$-moves allow us to switch between 
these POP decompositions,
and the axioms for the modular
functor make these consistent.
Here we show two such triangle consistency requirements
(they are reflections of each other):
\begin{center}
\includegraphics[width=0.25\columnwidth]{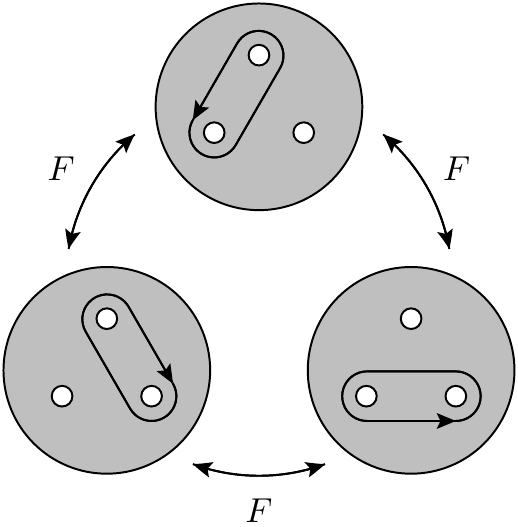}
\ \ \ \ \ \ \ \ \ \ \ \ \ 
\includegraphics[width=0.25\columnwidth]{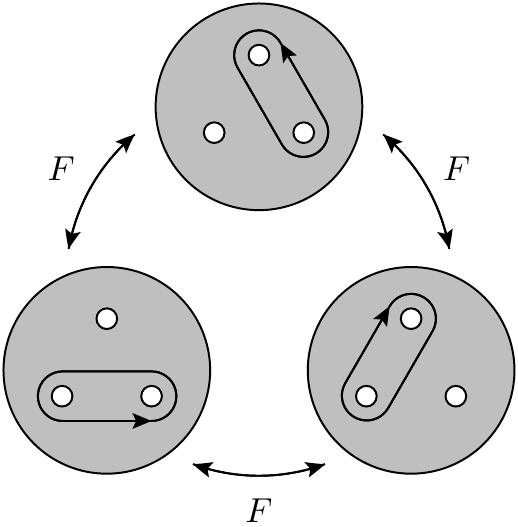}
\end{center}
Given a POP decomposition of a surface $N$
there are various curve diagrams on $N$ that
produce this decomposition from 
a standard POP decomposition, and there are certain $R$-moves
that will map between these.
Once again, these moves must be consistent, and
here we note the two ``hexagon equations'' 
corresponding to the above two triangles:
\begin{center}
\includegraphics[width=0.4\columnwidth]{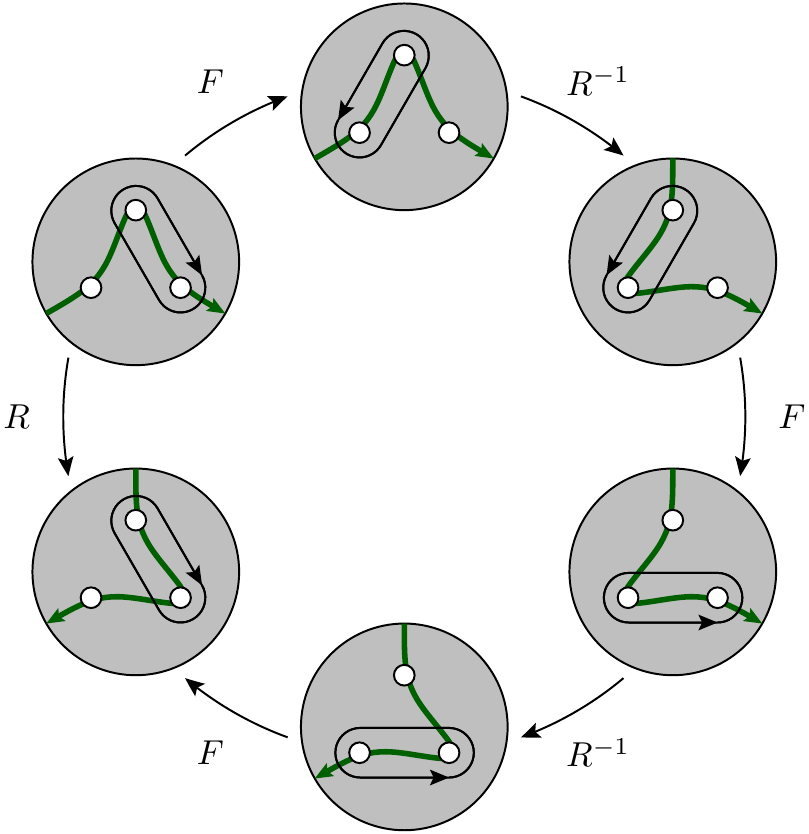}
\ \ \ \ \ \ \ \ \ \ \ \ 
\includegraphics[width=0.4\columnwidth]{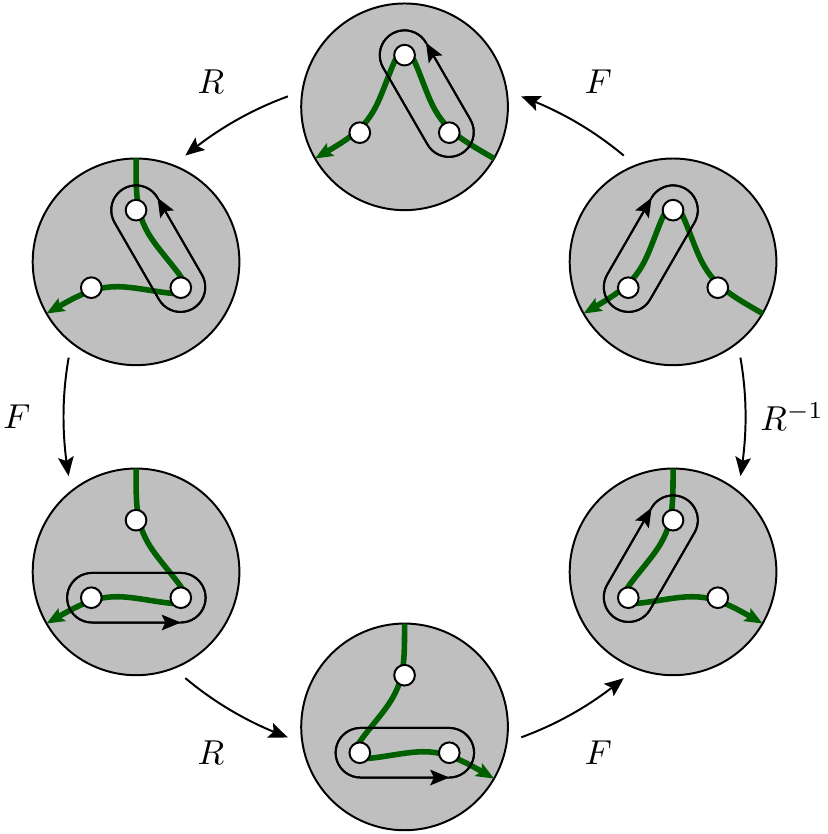}
\end{center}
Note that we have neglected to indicate the base points here.
Given a curve diagram, we can agree that base points
occur ``to the right'' of the image of the equator.
But in notating diagrams such as these, there is still
an ambiguity: the position of the base points should be the
same in each surface in the diagram.
If we try to correct for this post-hoc
by rotating individual holes,
we will then be correct
only up to possible Dehn twist(s) around each hole.
We can certainly track these twists if we wanted to,
but in the interests of simplicity we do not.
This introduces a global phase ambiguity into the
calculations.


The reason why we mention the pentagon and hexagon
equations is that these become important in an
algebraic description of the theory.
Because we have defined everything in terms of
a modular functor we get these equations ``for free''.


%
%

\subsection{Refactoring theorem}
In this section we specialize to considering planar surfaces only.
The theorem we are building towards shows that
the $R$-moves act transitively
on curve diagrams.
By transitive we mean that
given two curve diagrams $f$ and $f'$ on $N$ we can find
a sequence of $R$-moves that transform $f$ to $f'.$
The key idea is to consider the (directed) image of the equator under
these curve diagrams.
As mentioned previously, this image is sufficient to
define the entire diffeomorphism (up to isotopy) and so we are free
to work with just this image, or \emph{curve},
and we confuse the distinction between 
a curve diagram (a diffeomorphism)
and its curve.

The construction proceeds by considering adjacent
pairs of holes along $f'$ and then acting on the
portion of $f$ between these same two holes so that
they then become adjacent on the resulting curve.
Continuing this
process for each two adjacent holes of $f'$ will then
show a sequence of $R$-moves that sends $f$ to $f'.$

To this end, consider a sequence of holes $a_1,...,a_n$
appearing sequentially along $f$
such that $a_1$ and $a_n$ appear sequentially on $f'.$
We may need to apply some $Z$-moves to $f$ to ensure
that $a_1$ appears sequentially before $a_n$.
Now consider the case such that
along $f'$ between these two holes there is
no intersection with $f.$
We form a closed path $\xi$ by
following the $f'$ curve between $a_1$ and $a_n$ and then
following the $f$ curve in reverse from $a_n$ back to $a_1.$
(Note that to be completely rigorous here we would need to include segments
of path contained within boundary components.)
The resulting closed path bounds a disc.
If $\xi$ has clockwise orientation 
we apply the following sequence of $R$-moves to $f$:
$$
    R^{a_1a_2}_{b_1}, R^{a_1a_3}_{b_2}, ..., R^{a_1a_{n-1}}_{b_{n-1}}.
$$
Depicted here is the first such move:
\begin{center}
\includegraphics[]{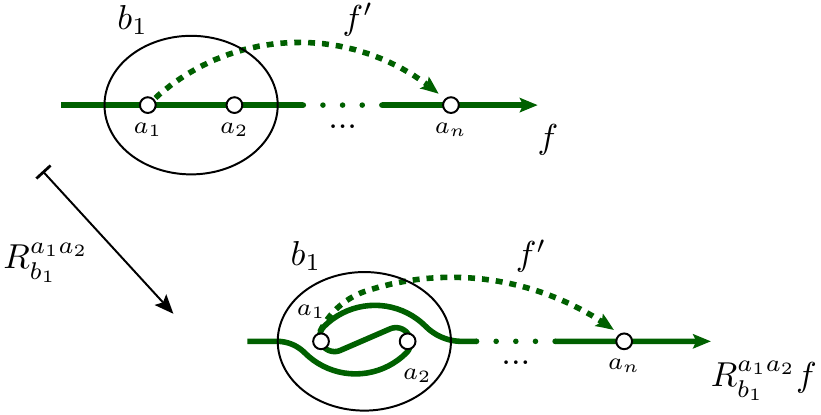}
\end{center}
After each of these $R$-moves the closed path
formed by 
following the $f'$ curve between $a_1$ and $a_n$ and then
following the $R$-moved $f$ curve back to $a_1$ will
traverse one less hole, and still bound a disc.
After all of the $R$-moves this path will only
touch $a_1$ and $a_n,$ and bound a disc.
Therefore, we have acted on the $f$ curve so that
the resulting curve has $a_1$ and $a_n$ adjacent
and the bounded disc gives an isotopy for that
segment of the curve.

When the closed curve $\xi$ has anti-clockwise orientation
we use the same sequence of $R$-moves but with $R$ replaced by $R^{-1}.$

Generalizing further,
when the $f'$ curve between $a_1$ and $a_n$ has (transverse) intersections
with $f$ we use every such intersection to indicate a switch
between using $R$ and $R^{-1}.$

We continue in this way moving backwards (from
head to tail) sequentially applying this procedure.




\begin{acknowledgments}

The author wishes to thank 
Andrew Doherty, Courtney Brell, Steven Flammia and Dominic Williamson
for helpful discussions and inspiration.

\end{acknowledgments}

%
%
\bibliography{refs2}

\end{document}